\documentclass
[preprint,prb,a4paper,amsfonts,amssymb,floats,titlepage,fleqn,footinbib,12pt,onecolumn,nobalancelastpage,showpacs]{revtex4}%
\usepackage{amsfonts}
\usepackage{amsmath}
\usepackage{amssymb}
\usepackage{graphicx}
\usepackage{longtable}%
\begin{document}
\author{M. ElMassalami, R.\ E. Rapp, F. A. B. Chaves and R. Moreno}
\affiliation{Instituto de Fisica, Universidade Federal do Rio de Janeiro, Caixa Postal
68528, 21945-972 Rio de Janeiro, Brazil}
\author{H. Takeya}
\affiliation{National Institute for Materials Science,1-2-1 Sengen, Tsukuba, Ibaraki
305-0047, Japan}
\author{B. Ouladdiaf }
\affiliation{Institut Laue-Langevin, B.P.156 ,38042 Grenoble Cedex 9, France}
\author{J. W. Lynn, Q. Huang}
\affiliation{NIST Center for Neutron Research, National Institute of Standards and
Technology, Gaithersburg, MD 20899-6102}
\author{R. S. Freitas and N. F. Oliveria Jr.}
\affiliation{Instituto de F\'{\i}sica., Universidade de S\~{a}o Paulo, Rua do Mat\~{a}o 187
Travessa R, Cidade Universit\'{a}ria, 05315-970 Sao Paulo, SP , Brasil }
\title{Synthesis and magnetic characterization of TmCo$_{2}$B$_{2}$C}
\date{\today{}}

\begin{abstract}
A new quaternary intermetallic borocarbide TmCo$_{2}$B$_{2}$C has been
synthesized via a rapid-quench of an arc-melted ingot. Elemental and
powder-diffraction analyses established its correct stoichiometry and
single-phase character. The crystal structure is isomorphous to that of
TmNi$_{2}$B$_{2}$C ($I4/mmm$) and is stable over the studied temperature
range. Above 7 K, the paramagnetic state follows the modified Curie-Weiss
behavior ($\chi=C/(T-\theta)+\chi_{0}$) wherein $\chi_{0}$=0.008(1) emu/mole
and the temperature-dependent term reflecting the paramagnetism of the Tm
subsystem: $\mu_{\text{eff}}=$7.6(2) $\mu_{\text{B}}$ [in agreement with the
expected value for a free Tm$^{3+}$ ion] and $\theta$ = -4.5(3) K. Long range
ferromagnetic order of the Tm sublattice is observed to develop around
$\mathrm{\sim}$1 K. No superconductivity is detected in TmCo$_{2}$B$_{2}$C
down to 20 mK, a feature which is consistent with the general trend in the
RCo$_{2}$B$_{2}$C series. Finally, the influence of the rapid-quench process
on the magnetism (and superconductivity) of TmNi$_{2}$B$_{2}$C will be
discussed and compared to that of TmCo$_{2}$B$_{2}$C.

\end{abstract}

\maketitle

\section{Introduction}

The magnetic ordering temperature of a Tm$^{3+}$ sublattice in a thulium-based
intermetallic compound is, in general, lower than that of its isomorphous
heavy rare-earth-based counterparts of the same series.\cite{Coqblin-book} One
usually attributes such a lowering to a smaller de Gennes factor,
$(g-1)^{2}J(J+1)$, assuming that parameters such as electronic structure are
not modified across the isomorphous series. On the other hand, for Tm-based
compounds that belong to a family of different series, electronic properties
(such as the position of the Fermi level within the\ density of states curve,
the derivative of the density of states, spin fluctuations, ... etc.) do
change and such a change can modify significantly the type, character, and
critical point of the magnetic order. Such features are well documented in,
say, the $RM_{2}$ family ($R=$ magnetic rare earth, $M=3d$ transition
metal):\cite{Bloch-Lemaire70-RCo2,Bloch75-RCo2,Cyrot79-electr-cal,Inoue88} for
the \textrm{Tm}$M_{2}$ compounds, the critical temperature of each member
depends critically on the type of $M$ but, nevertheless for each $RM_{2}$
series, it is always lower than that of the corresponding heavy $R$-isomorphs.

Another illustration which is of relevance to this work is the case of the
heavier members of the borocarbide family $RM_{2}$\textrm{B}$_{2}$\textrm{C.}
For\ the $R\mathrm{Ni}_{2}$\textrm{B}$_{2}$\textrm{C} series, the Neel
temperatures ($T_{N}$) scale reasonably well with the de Gennes
factor;\cite{Eisaki94-RNi2B2C,Cho96-RNi2B2C-deGeness,Lynn97-RNi2B2C-ND-mag-crys-structure}
moreover, both $T_{N}$\textrm{ }(1.5 K) and the de Gennes factor (1.17) of
\textrm{TmNi}$_{2}$\textrm{B}$_{2}$\textrm{C} are the lowest. In contrast, the
critical points $T_{C}$ of the isomorphous series $R$\textrm{Co}$_{2}%
$\textrm{B}$_{2}$\textrm{C} do not show this scaling. Furthermore, $T_{C}$ of
\textrm{TmCo}$_{2}$\textrm{B}$_{2}$\textrm{C} has not, so far, been reported,
though based on an extrapolation of $T_{C}$ within the phase diagram of
$R$\textrm{Co}$_{2}$\textrm{B}$_{2}$\textrm{C }($T_{C}$ versus the de-Gennes
factor), it is expected to be 2 K.\cite{00-RCo2B2C} But this would contradict
the observed trend that $T_{C}$ 's\ of the heavier members of $R$%
\textrm{Co}$_{2}$\textrm{B}$_{2}$\textrm{C} are not higher than those of the
corresponding $R$\textrm{Ni}$_{2}$\textrm{B}$_{2}$\textrm{C}. This question
has not been addressed before due to the difficulties in the synthesis and
stabilization of a single-phase \textrm{TmCo}$_{2}$\textrm{B}$_{2}$\textrm{C
}compound. This work reports on the successful preparation as well as the
structural and physical characterization of single-phase \textrm{TmCo}$_{2}%
$\textrm{B}$_{2}$\textrm{C} samples. The obtained results confirm that indeed
its $T_{C}$ is not higher than that of \textrm{TmNi}$_{2}$\textrm{B}$_{2}%
$\textrm{C}.

Two findings are of special interest to the understanding of the
superconductivity, magnetism, and their interplay in the borocarbide family:
First, no sign of superconductivity is observed down to 20 mK; such a result,
together with the nonsuperconductivity of \textrm{YCo}$_{2}$\textrm{B}$_{2}%
$\textrm{C},\cite{00-RCo2B2C,01-PrDyCo2B2C} indicate that the quench of the
superconductivity in $R$\textrm{Co}$_{2}$\textrm{B}$_{2}$\textrm{C} is related
to the unfavorable spin fluctuation process (for the particular case of
\textrm{TmCo}$_{2}$\textrm{B}$_{2}$\textrm{C}, there is, in addition, the
pair-braking ferromagnetic, FM, order). Secondly, the long-range order of the
localized 4$f$ moments of \textrm{TmCo}$_{2}$\textrm{B}$_{2}$\textrm{C} does
not appear to be a spin density wave as in the case of \textrm{TmNi}$_{2}%
$\textrm{B}$_{2}$\textrm{C}; rather, we argue that a FM state is consistent
with the results of the magnetization, specific heat, and neutron diffraction.
These findings will be demonstrated and discussed in \S \ III. Before that we
describe, in \S \ II, the synthesis process and show how the stoichiometry and
the single-phase character\ of the rapid-quenched ($rq$) \textrm{TmCo}$_{2}%
$\textrm{B}$_{2}$\textrm{C} and \textrm{TmNi}$_{2}$\textrm{B}$_{2}$\textrm{C}
samples were verified. An evaluation of the influence of the rate of the $rq$
process on the studied physical properties of \textrm{TmCo}$_{2}$%
\textrm{B}$_{2}$\textrm{C} (\S \ II)\ together with a detailed comparison of
the properties of a normal-prepared, non-quenched ($nq$) \textrm{TmNi}$_{2}%
$\textrm{B}$_{2}$\textrm{C} with those of a $rq$ \textrm{TmNi}$_{2}$%
\textrm{B}$_{2}$\textrm{C} (\S \ II) would allow us to extract the dominant
influences of this \textit{rq} process.

\section{Experiment}

\subsection{\textit{rq}-\textrm{TmCo}$_{2}$\textrm{B}$_{2}$\textrm{C}}

A starting sample with a stoichiometry of \textrm{TmCo}$_{2}$\textrm{B}$_{2}%
$\textrm{C} was prepared by a conventional arc-melt procedure under a highly
pure (99.999\%) argon atmosphere. The obtained product, when tested with X-ray
diffraction, showed a multi-phase pattern. This same product, shaped into
balls with diameters around 3--5 mm, was remelted and directly rapid-quenched
by hitting it with a copper hammer.\cite{Takeya04-NbB2} The final product is a
single-phase \textrm{TmCo}$_{2}$\textrm{B}$_{2}$\textrm{C} in the form of thin
flakes (50 $\sim$ 100 $\mu$m thickness). Once obtained, the sample is stable
in air and does not need special care during the handling or storing process,
at least within an interval of months.\begin{table}[t]
\caption{The molar ratios of the $^{11}$B-enriched, rapid-quenched
\textrm{TmCo}$_{2}$\textrm{B}$_{2}$\textrm{C} and \textrm{TmNi}$_{2}%
$\textrm{B}$_{2}$\textrm{C }samples. The calculations are based on the weight
percentage data (normalized to the Tm content) and using the following atomic
weights: Tm=168.93421(3), Co=58.93320(1), Ni=58.6934, $^{11}$B=10.995 (99.5\%
enriched from Eagle Pitcher Ind. Inc.), B =10.811(5), and C =12.011(1). The
room-temperature cell parameters of \textrm{TmCo}$_{2}$\textrm{B}$_{2}%
$\textrm{C}, as obtained from XRD analysis (see Fig. \ref{Fig.1}), are $a$=
3.473(1)$\operatorname{\mathring{A}},c$= 10.647(4)$\operatorname{\mathring{A}%
}$. For \textrm{TmNi}$_{2}$\textrm{B}$_{2}$\textrm{C}, see Table II.}%
\label{Tab.I}%
\begin{tabular}
[c]{lllllll}\hline\hline
compound & ratio & Tm & Co & Ni & B & C\\\hline
TmCo$_{2}$B$_{2}$C & \% & 52.2(1) & 36.3(1) & - & 6.69(1) & 3.81(1)\\
& molar & 1 & 1.993(5) & - & 1.97(3) & 1.03(3)\\
TmNi$_{2}$B$_{2}$C & \% & 52.8(1) & - & 36.0(1) & 6.39(1) & 3.77(1)\\
& molar & 1 & - & 1.962(6) & 1.86(3) & 1.00(3)\\\hline\hline
\end{tabular}
\end{table}

\begin{figure}
[th]
\begin{center}
\includegraphics[
height=4.2298in,
width=3.0096in
]%
{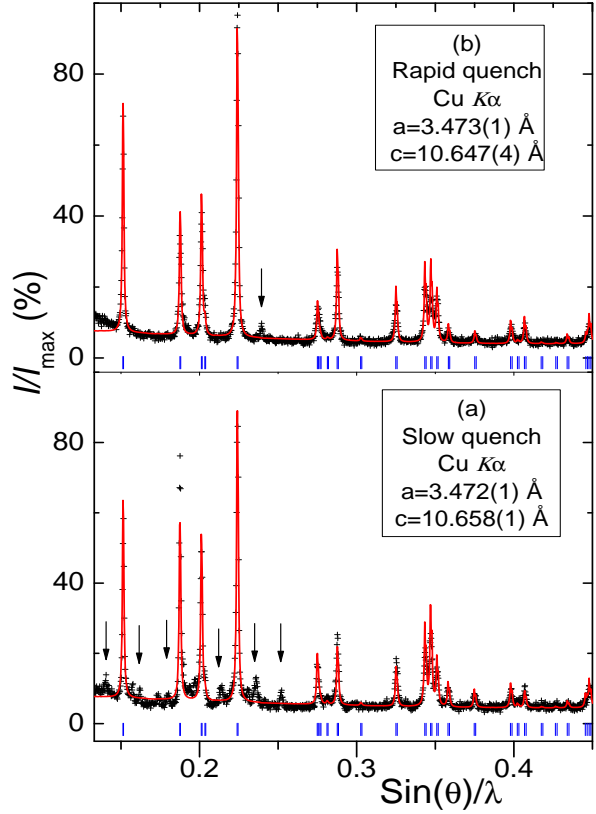}%
\caption{(Color online) Room temperature X-ray diffractograms (Cu $K_{\alpha}%
$) of (a) slow-quenched and (b) rapid-quenched \textrm{TmCo}$_{2}$%
\textrm{B}$_{2}$\textrm{C}. Rietveld analysis (solid lines) shows that the
lattice parameters are hardly affected by the variation of the quench rate.
Nevertheless, the impurity content are noticeably suppressed. On comparison
with the neutron diffractograms (see later), allowance should be made for the
difference in the $f$-factors.}%
\label{Fig.1}%
\end{center}
\end{figure}

In contrast to the $R$\textrm{Ni}$_{2}$\textrm{B}$_{2}$\textrm{C}
series\cite{Siegrist94a-RNi2B2C,Siegrist94b-RNi2B2C} or even to other
$R$\textrm{Co}$_{2}$\textrm{B}$_{2}$\textrm{C }members,\cite{00-RCo2B2C} the
$rq$ process is essential for the synthesis of \textrm{TmCo}$_{2}$%
\textrm{B}$_{2}$\textrm{C}. Fig. \ref{Fig.1} shows the diffractograms of two
samples of \textrm{TmCo}$_{2}$\textrm{B}$_{2}$\textrm{C} prepared with
different quenching rates (the one with a faster quench rate is enriched
with\ $^{11}$\textrm{B}). These diffractograms together with results from
various measuring techniques reveal that the $rq$ process drastically
diminishes the amount of impurity phases and as such leads to noticeable
changes in the physical properties (see below, in particular\ Fig.
\ref{Fig.4}). All measurements reported below were carried out on the same
as-prepared, rapid-quenched, $^{11}$B-enriched batch: conventional annealing
(often followed for the borocarbides samples) leads to a surge of additional contaminations.%

The elemental analysis of both \textit{rq}-\textrm{TmCo}$_{2}$\textrm{B}$_{2}%
$\textrm{C} and \textit{rq}-\textrm{TmNi}$_{2}$\textrm{B}$_{2}$\textrm{C }is
shown in Table \ref{Tab.I}. The content of Tm, Co, and B were determined by
the Induction Coupled Plasma analysis while that of C by the process of carbon
combustion followed by Infrared absorption analysis. Table \ref{Tab.I} shows
also the cell parameters as obtained from room-temperature X-ray diffraction
analysis [Fig. \ref{Fig.1} (b)]. As can be verified, these results (in
particular that of B and C) do confirm the correct stoichiometry of the
\textrm{TmCo}$_{2}$\textrm{B}$_{2}$\textrm{C} samples.

The crystal structure was studied by X-ray diffraction (room-temperature) and
neutron diffraction (0.36 $<T<$ 30 K). Physical characterizations were carried
out with the following techniques: magnetization [$M(T,H),$ 0.5 $<T<$ 30 K,
$H$%
$<$
150 kOe], $dc$ susceptibility\ [$\chi_{dc}(T)=M/H$, 0.5 $<T<$ 300 K, $H\leq$10
kOe], $ac$ susceptibility\ [$\chi_{ac}(T)$, 0.02 $<T<$ 20 K, $f$=200 $\sim
$1000 Hz, $h_{ac}$%
$<$
10 Oe], and specific heat [$C(T)$, semi-adiabatic, 0.1 $<T<$ 15 K].
Neutron-diffraction was carried out at two different sites: (i) the Institut
Laue-Langevin (ILL, Grenoble, France) using the D2B diffractometer (3 $\leq
T\leq$ 30 K) with a wavelength of $\lambda=$1.595 $%
\operatorname{\mathring{A}}%
$, and (ii) the National Institute of Standards and Technology (NIST,
Gaithersburg, USA) using the high-resolution powder diffractometer ($\lambda
=$2.0787 $%
\operatorname{\mathring{A}}%
$, $T$=0.36 K) and the BT-9 triple-axis instrument with a pyrolytic graphite
monochromator and filter ($\lambda=$2.359 $%
\operatorname{\mathring{A}}%
$, 0.48 $\leq T\leq$ 4.5 K). Data were also collected on the new BT-7 triple
axis spectrometer using the position sensitive detector in diffraction mode.
For these neutron measurements a sample weighing 0.66 grams was used.

\subsection{\textit{rq}-\textrm{TmNi}$_{2}$\textrm{B}$_{2}$\textrm{C}}

\begin{table}[t]
\caption{Comparison between the room-temperature cell parameters ($a,$ $c$),
the $z$ parameter indicating the position of the B atom (the positions of the
other atoms are fixed by symmetry), and the paramagnetic Curie-Weiss
parameters (figures not shown) of the rapid-quenched and non-quenched sample
of \textrm{TmNi}$_{2}$\textrm{B}$_{2}$\textrm{C}. For the non--quenched
sample, the cell parameters are taken from Lynn\textit{ et al.}
\cite{Lynn97-RNi2B2C-ND-mag-crys-structure} while the CW parameters are from
Cho\textit{ et al.}\cite{Cho95-Tm-superconductivity}}%
\label{Tab.II}%
\begin{tabular}
[c]{llllll}\cline{1-2}\cline{1-3}\cline{3-6}\cline{4-6}%
TmNi$_{2}$B$_{2}$C & a ($\operatorname{\mathring{A}}$) & c
($\operatorname{\mathring{A}}$) & $z$ & $p_{eff}$($\mu_{B}$) & $\theta
$(K)\\\hline
rapid-quenched & 3.4857(9) & 10.5609(27) & 0.3519 & 7.6(1) & --7.5(2)\\
non-quenched & 3.4866(2) & 10.5860(5) & 0.3598(2) & 7.54(2) &
-11.6(4)\\\hline\hline
\end{tabular}
\end{table}%
\begin{figure}
[th]
\begin{center}
\includegraphics[
height=3.4809in,
width=2.5573in
]%
{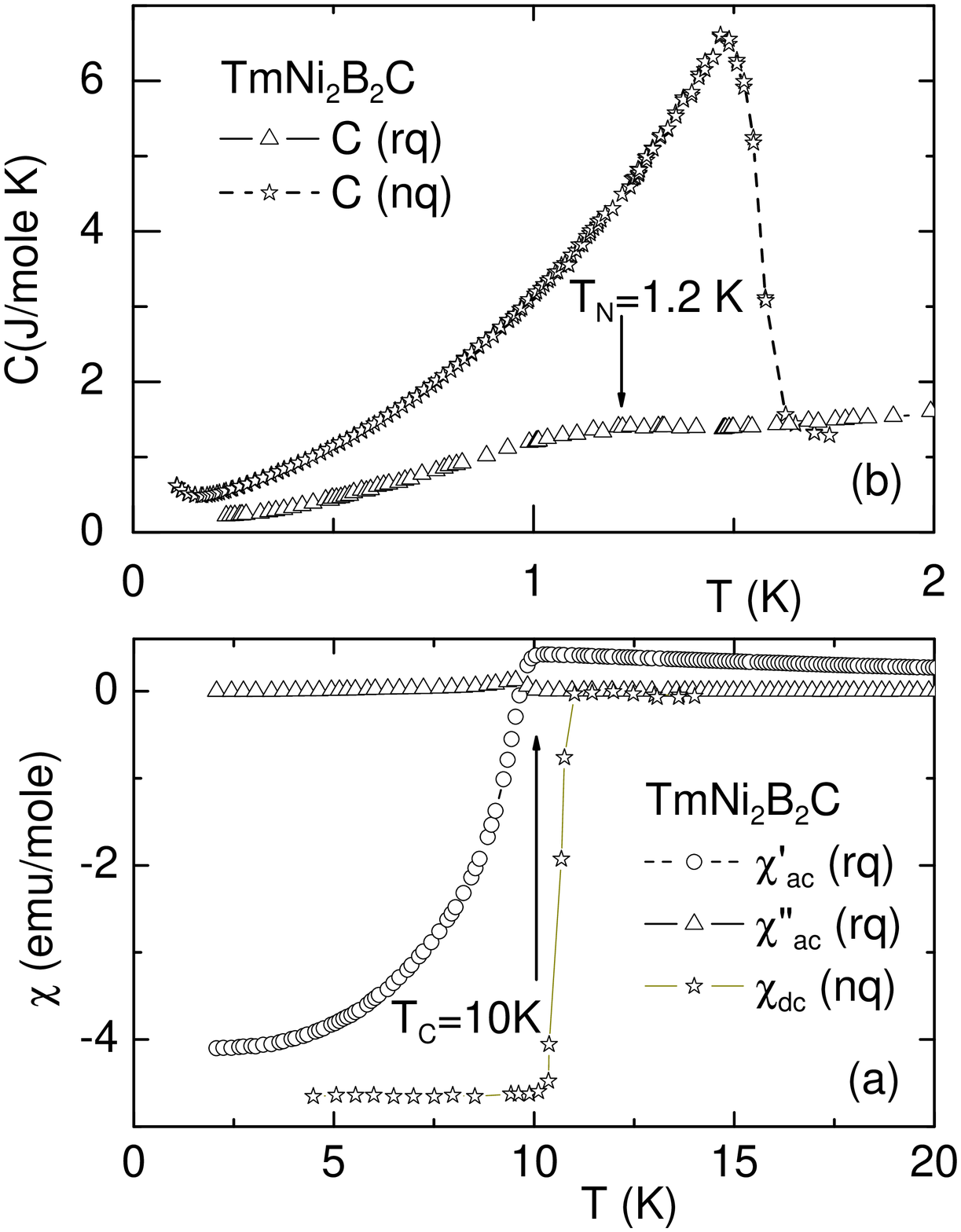}%
\caption{(a) Low-temperature, low-field $ac$ ($\chi_{ac}$) and $dc$
($\chi_{dc}$) susceptibilities of \textrm{TmNi}$_{2}$\textrm{B}$_{2}%
$\textrm{C}. $\chi_{ac}$ ($H_{ac}$= 1 Oe, f = 500 Hz) is measured on $rq$
sample while $\chi_{dc}$ ($H_{dc}=10$ Oe) is measured by Cho \textit{et al.}
\cite{Cho95-Tm-superconductivity} on $nq$ sample. (b) The total specific heats
of $rq$ and $nq$ \textrm{TmNi}$_{2}$\textrm{B}$_{2}$\textrm{C} samples. The
later is taken from Movshovich \textit{et al.}\cite{Movshovich94-TmNi2B2C}
Note the large broadening of the superconducting and magnetic transitions of
the $rq$ samples.}%
\label{Fig.2}%
\end{center}
\end{figure}
Intended as a helpful guide in the evaluation of how much the properties of
TmCo$_{2}$B$_{2}$C are being influenced by the $rq$ process, we investigated
also the physical properties of a $rq$ \textrm{TmNi}$_{2}$\textrm{B}$_{2}%
$\textrm{C} sample and compare its results with those of the reported $nq$
\textrm{TmNi}$_{2}$\textrm{B}$_{2}$\textrm{C} sample. It is recalled that a
$nq$ \textrm{TmNi}$_{2}$\textrm{B}$_{2}$\textrm{C} sample shows typical
second-order-type superconducting and magnetic phase transitions occurring,
respectively, at $T_{c}$ =11 K and $T_{N}=$1.5
K.\cite{Cho95-Tm-superconductivity,Movshovich94-TmNi2B2C} We synthesized a
$rq$ \textrm{TmNi}$_{2}$\textrm{B}$_{2}$\textrm{C} via an identical
preparation procedure as the one used for \textrm{TmCo}$_{2}$\textrm{B}$_{2}%
$\textrm{C}; the resulting sample has been studied by room-temperature XRD,
specific heat, and magnetization measurements. The cell parameters and the
paramagnetic CW behavior (see Table \ref{Tab.II}) reflect slight modifications
which may be related to the $rq$ process and a slight depletion in B content.
It is noted that the $rq$ process does not induce any noticeable moment on the Ni-sublattice.

Figure \ref{Fig.2} shows that both the magnetic and superconducting order of
\textrm{TmNi}$_{2}$\textrm{B}$_{2}$\textrm{C} do survive the $rq$ process
(only that $T_{N}$ and $T_{c}$ are lowered to 1.2 K and 10 K, respectively)
suggesting that the intrinsic forces that govern these cooperative phenomena
are much stronger than the quench-induced disordering effects. Nevertheless,
the $rq$ process has modified substantially the structure of the curves within
the transition region: the width of the superconducting phase transition [Fig.
\ref{Fig.2}(a)] is substantially increased while the height (width) of the
magnetic phase transition is strongly reduced (increased) to the extent that
the magnetic transition is manifested as a weak, broadened event; the
$\lambda$-type magnetic transition of the $nq$ sample is being transformed by
the $rq$ process into an almost featureless event; in this case the phase
change, in contrast to the $nq$ case, is evidenced in the features of the
first-order derivative: thus the \textit{rq} process may modify the character
and type of the magnetic phase transition. As we observed no hysteresis
effects, the assignment of a first-order character to this phase transition is
not unambiguous. At any rate, it can be safely concluded that the
superconducting (and to some extent the magnetic) features of both the\textit{
rq}- and \textit{nq}-\textrm{TmNi}$_{2}$\textrm{B}$_{2}$\textrm{C}, except for
the region neighboring the critical temperature, are practically similar. It
is then concluded that the tetragonal \textrm{TmNi}$_{2}$\textrm{B}$_{2}%
$\textrm{C} structure, in sharp contrast with \textrm{TmCo}$_{2}$%
\textrm{B}$_{2}$\textrm{C}, is a low-temperature phase. Nevertheless, the
observation that both isomorphs have almost the same structural-chemical
properties suggests that the overall effect of the $rq$ process on the
physical properties of \textrm{TmCo}$_{2}$\textrm{B}$_{2}$\textrm{C} would not
differ much from the trend observed in \textrm{TmNi}$_{2}$\textrm{B}$_{2}%
$\textrm{C}; in particular as that the $rq$ process did not modify drastically
the the intrinsic physical properties of \textrm{TmNi}$_{2}$\textrm{B}$_{2}%
$\textrm{C}, then both features of \textrm{TmCo}$_{2}$\textrm{B}$_{2}%
$\textrm{C} (the nonsuperconductivity and the FM mode, see below) are
considered to be intrinsic properties that are not strongly influenced by the
\textit{rq} process.

\section{Results and Discussion}

\begin{figure}
[th]
\begin{center}
\includegraphics[
height=3.3676in,
width=2.5157in
]%
{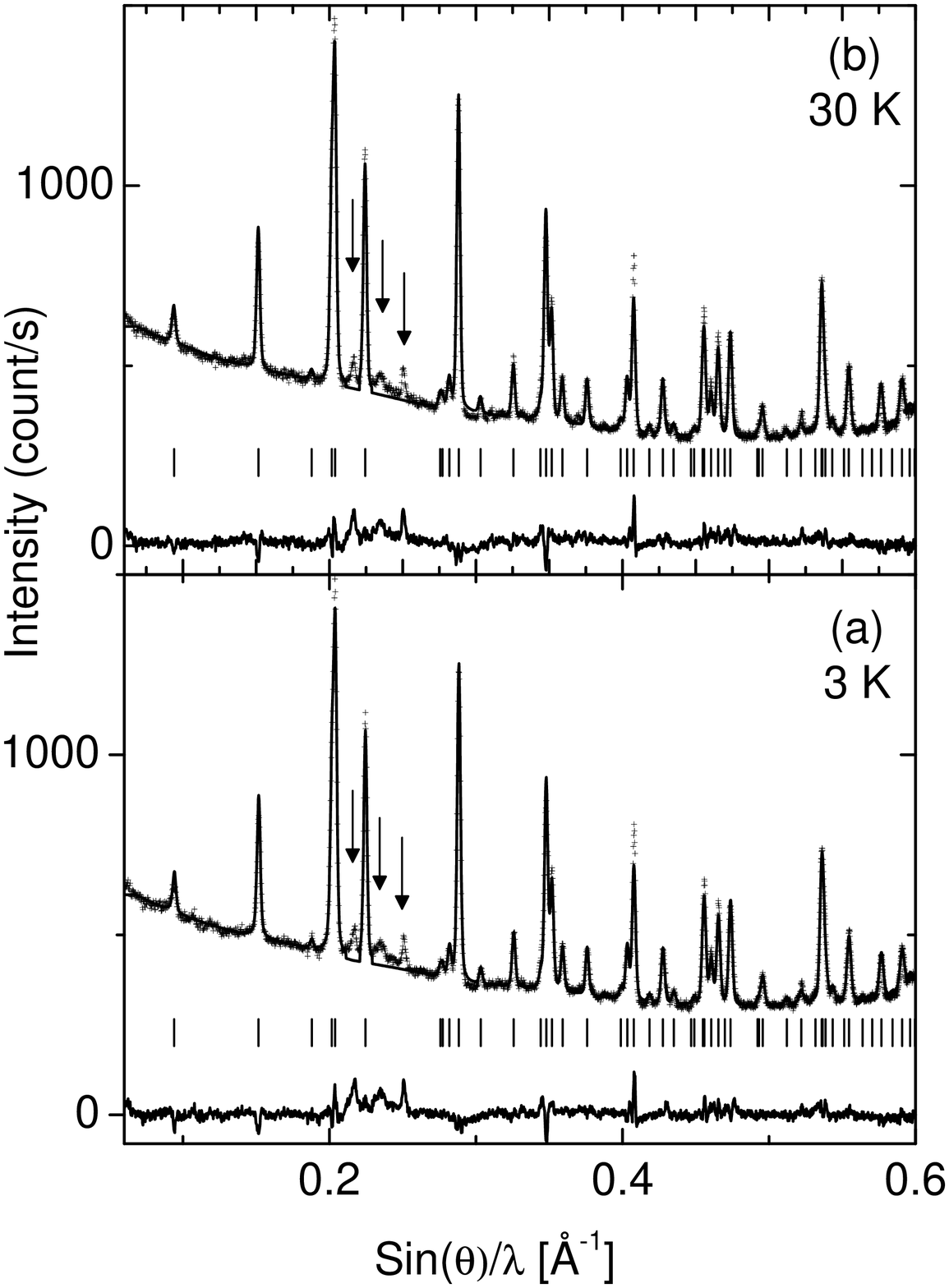}%
\caption{Neutron powder diffractograms of \textrm{TmCo}$_{2}$\textrm{B}$_{2}%
$\textrm{C,} measured on D2B (ILL) diffractometer at (a) 3 K, and (b) 30 K.
Note the very weak impurity lines at $\sin(\theta)/\lambda$=0.217, 0.234, and
0.251 $\operatorname{\mathring{A}}^{-1}$ (see Fig. \ref{Fig.1} ) The intensity
of the strongest impurity\ line relative to that of the main phase amounts
only to 4\%.}%
\label{Fig.3}%
\end{center}
\end{figure}

The analysis of the powder diffractograms (Fig. \ref{Fig.3} and Table
\ref{Tab.III}) indicates that \textrm{TmCo}$_{2}$\textrm{B}$_{2}$\textrm{C
}crystallizes\textrm{ }in the tetragonal \textrm{LuNi}$_{2}$\textrm{B}$_{2}%
$\textrm{C}-type structure which, within the experimental accuracy, is stable
over the range 0.36 $\leq T\leq$ 300K. In comparison with isomorphous
\textrm{TmNi}$_{2}$\textrm{B}$_{2}$\textrm{C}, the introduction of Co reduces
the $a$-parameter and elongates the $c$-axis length but it induces no drastic
shift in the B $z$-parameter (no particular significance is attributed to the
observation that the $z$-parameters of both $rq$ samples are 2\% different
from the $nq$ \textrm{TmNi}$_{2}$\textrm{B}$_{2}$\textrm{C}). It is reassuring
that the obtained cell parameters, when plotted together with those of the
other $R$\textrm{Co}$_{2}$\textrm{B}$_{2}$\textrm{C} compounds, evolve
linearly with the effective metallic radius of the $R$ atom (see Fig. 2 of
Ref. 9\nocite{00-RCo2B2C}).%
\begin{table}[t]
\caption{Cell parameters ($a,c$) and the B-atom $z$-parameter of
\textrm{TmCo}$_{2}$\textrm{B}$_{2}$\textrm{C} at selected temperatures. The
experimental diffractograms and the theoretically calculated patterns are
shown in Figs. \ref{Fig.3} and \ref{Fig.9}. In the Rietveld analysis of the
diffractograms, the thermal parameters were found to be the same as those of
\textrm{TmNi}$_{2}$\textrm{B}$_{2}$\textrm{C} given in Lynn \textit{et al.}
\cite{Lynn97-RNi2B2C-ND-mag-crys-structure} while the occupation parameters
gave similar values as the ones obtained from elemental analysis (see Table I)
}%
\label{Tab.III}%
\begin{tabular}
[c]{llll}\hline\hline
$T$ (K) & a ($\operatorname{\mathring{A}}$) & c ($\operatorname{\mathring{A}}%
$) & $z$\\\hline
30 & 3.4684(2) & 10.6411(6) & 0.3597(2)\\
3 & 3.4681(2) & 10.6413(7) & 0.3596(2)\\
0.36 & 3.4689(2) & 10.6438(9) & 0.3587(3)\\\hline\hline
\end{tabular}
\end{table}

\subsection{Magnetic susceptibilities (\textit{ac} and \textit{dc})}%

\begin{figure}
[th]
\begin{center}
\includegraphics[
height=3.4627in,
width=2.5244in
]%
{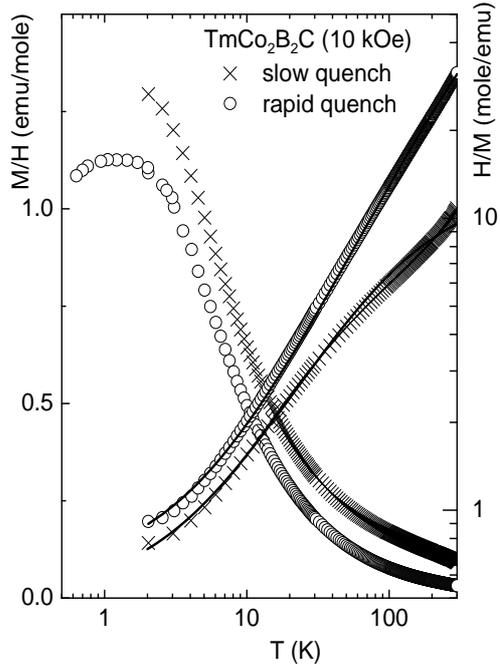}%
\caption{Temperature-dependent $\chi_{dc}=M/H$ curves at $H$=10 kOe of
\textrm{TmCo}$_{2}$\textrm{B}$_{2}$\textrm{C}. The temperature axis is
logarithmic so as to emphasize the low-temperature part. The influence of the
rate of the quenching process is demonstrated by comparison with a
slow-quenched sample (see Fig. \ref{Fig.1}). The reciprocal $\chi_{dc}$ (right
ordinate, logarithmic) for the $rq$ sample (represented by circles) is well
described by the modified CW behavior which is shown as solid lines (see
text).}%
\label{Fig.4}%
\end{center}
\end{figure}
%

\begin{figure}
[th]
\begin{center}
\includegraphics[
height=3.6383in,
width=2.5547in
]%
{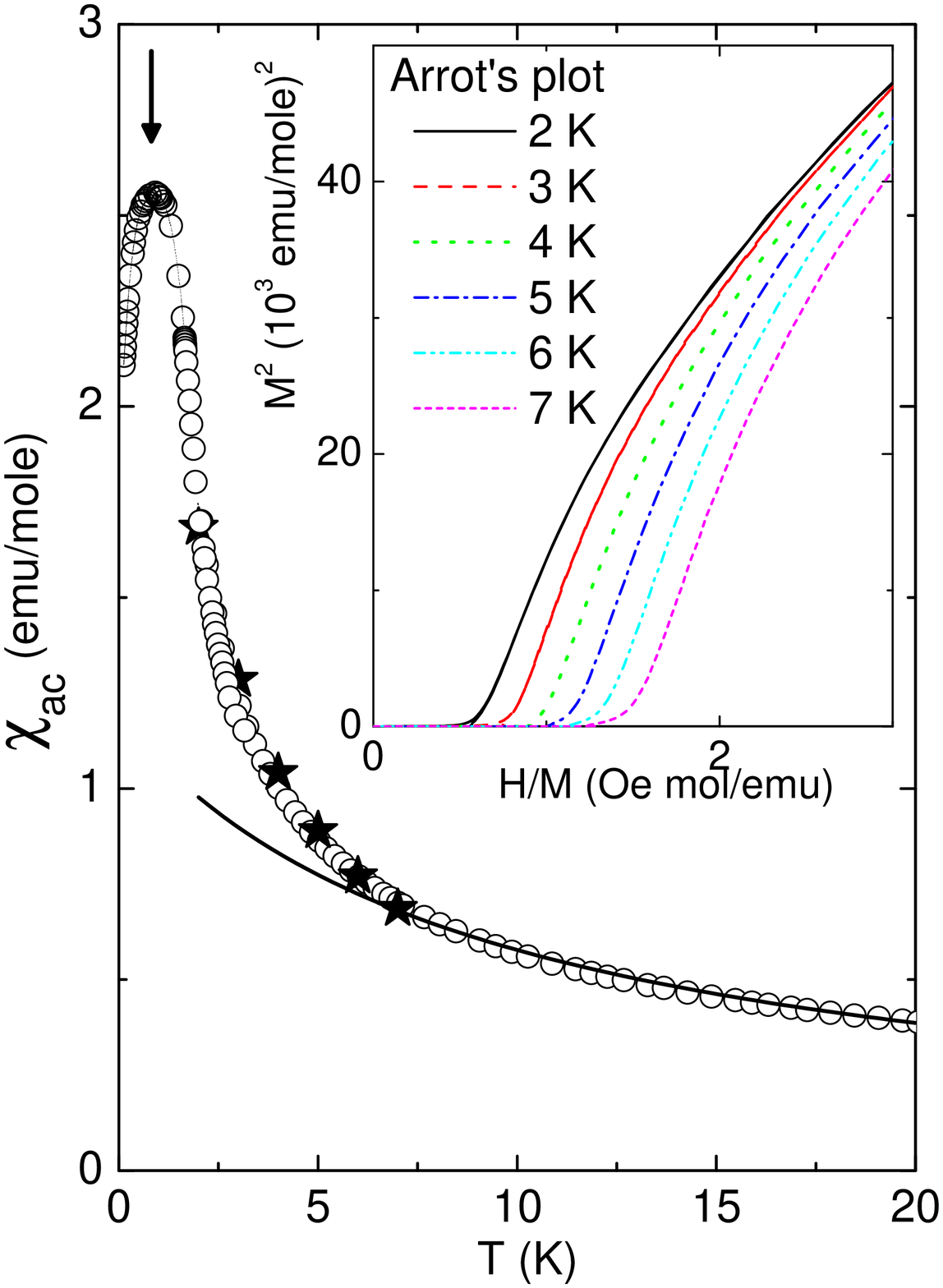}%
\caption{(Color online) Thermal evolution of the zero-field $\chi
_{ac}(T)=\partial M/\partial H$ of \ the \textit{rq} \textrm{TmCo}$_{2}%
$\textrm{B}$_{2}$\textrm{C}. The solid line represents a fit to the same
modified CW relation as in Fig. 3. The stars represent ($M/H)_{H\rightarrow
\text{0}}$ extrapolated from the Arrott\'{}s plots which are shown in the
inset: the agreement indicates that the influence of the magnetic impurities
within this range of temperatures is negligible.}%
\label{Fig.5}%
\end{center}
\end{figure}

Thermal evolution of the isofield $dc$ susceptibilities of the $rq$
\textrm{TmCo}$_{2}$\textrm{B}$_{2}$\textrm{C}, $\chi_{dc}=M/H$ (Fig.
\ref{Fig.4}),\textrm{ }follows faithfully, down to 7 K, the modified
Curie-Weiss (CW) behavior $\chi_{dc}(T)=C/(T-\theta)+\chi_{0}$, wherein
$\mu_{eff}=$7.6(2) $\mu_{\text{B}}$, $\theta=$ -4.5(3) K, and the
temperature-independent paramagnetic (TIP) contribution $\chi_{\text{0}}%
=$0.008(1) emu/mole. Similar high-temperature, modified CW paramagnetism is
also manifested in the zero-field $\chi_{ac}(T>$7 K$)=\partial M/\partial H$
curve (Fig. \ref{Fig.5}). $\chi_{dc}(T)$ of a slow-quenched (\textit{sq})
\textrm{TmCo}$_{2}$\textrm{B}$_{2}$\textrm{C} (shown also in Fig.
\ref{Fig.4})\ is higher in value and is not well described by the modified CW
law. Based on the structural analysis, the difference among these two curves
is related to the content of the contamination which, in turn, is governed by
the speed of the quenching rate (see Fig. \ref{Fig.1}): a faster rate leads to
a reduction in contamination and, consequently, to a lowering in the magnetic
susceptibility, in particular $\chi_{\text{0}}$. It is expected that on the
limit of a faster quench, there would be no impurity's contribution and
$\chi_{\text{0}}$ would be much smaller but nonvanishing since $\chi
_{\text{0}}$ of \textrm{YCo}$_{2}$\textrm{B}$_{2}$\textrm{C} is
nonzero.\cite{04-Pr(CoNi)2B2C} Such a limit $\chi_{\text{0}}$ may be due to a
van Vleck-type contribution or an exchange-enhanced Pauli susceptibility. The
admixture of high-lying orbitals into the ground state of the \textrm{Tm}%
$^{3+}$ ion can be ruled out since no such effect is observed in
\textrm{TmNi}$_{2}$\textrm{B}$_{2}$\textrm{C} [Ref.
14\nocite{Cho95-Tm-superconductivity}] (not even in the \textit{rq}-samples,
see Table \ref{Tab.II}). But such TIP is common in the cubic Laves
$R$\textrm{Co}$_{2}$ phases\cite{Bloch-Lemaire70-RCo2,Bloch75-RCo2} wherein
$\chi_{\text{0}}\approx$0.004 emu/mole. Though the above arguments as well as
the similar argument on \textrm{YCo}$_{2}$\textrm{B}$_{2}$\textrm{C} [Ref.
16\nocite{04-Pr(CoNi)2B2C}] suggest that such an additional $\chi_{\text{0}}$
may be related to the modification in the electronic structure of the
3\textit{d} subband (as a result of the introduction of the Co atoms), however
a final statement on this $\chi_{\text{0}}$ should wait for a further analysis.

As the temperature is lowered towards and below the liquid helium point, both
$\chi_{dc}(T)$ (Fig. \ref{Fig.4}) and $\chi_{ac}(T)$ (Fig. \ref{Fig.5})
manifest a rapid increase, going through a peak, and afterwards dropping
weakly downwards. Considering that $\chi_{ac}(T)$ has an accentuated peak with
a maximum at 0.8(1) K and that at this point $\chi_{dc}(T,$ 10 kOe$)$ starts
to decrease, such a point is taken to indicate a magnetic phase transition
(see below).
\begin{figure}
[th]
\begin{center}
\includegraphics[
height=3.4843in,
width=2.5192in
]%
{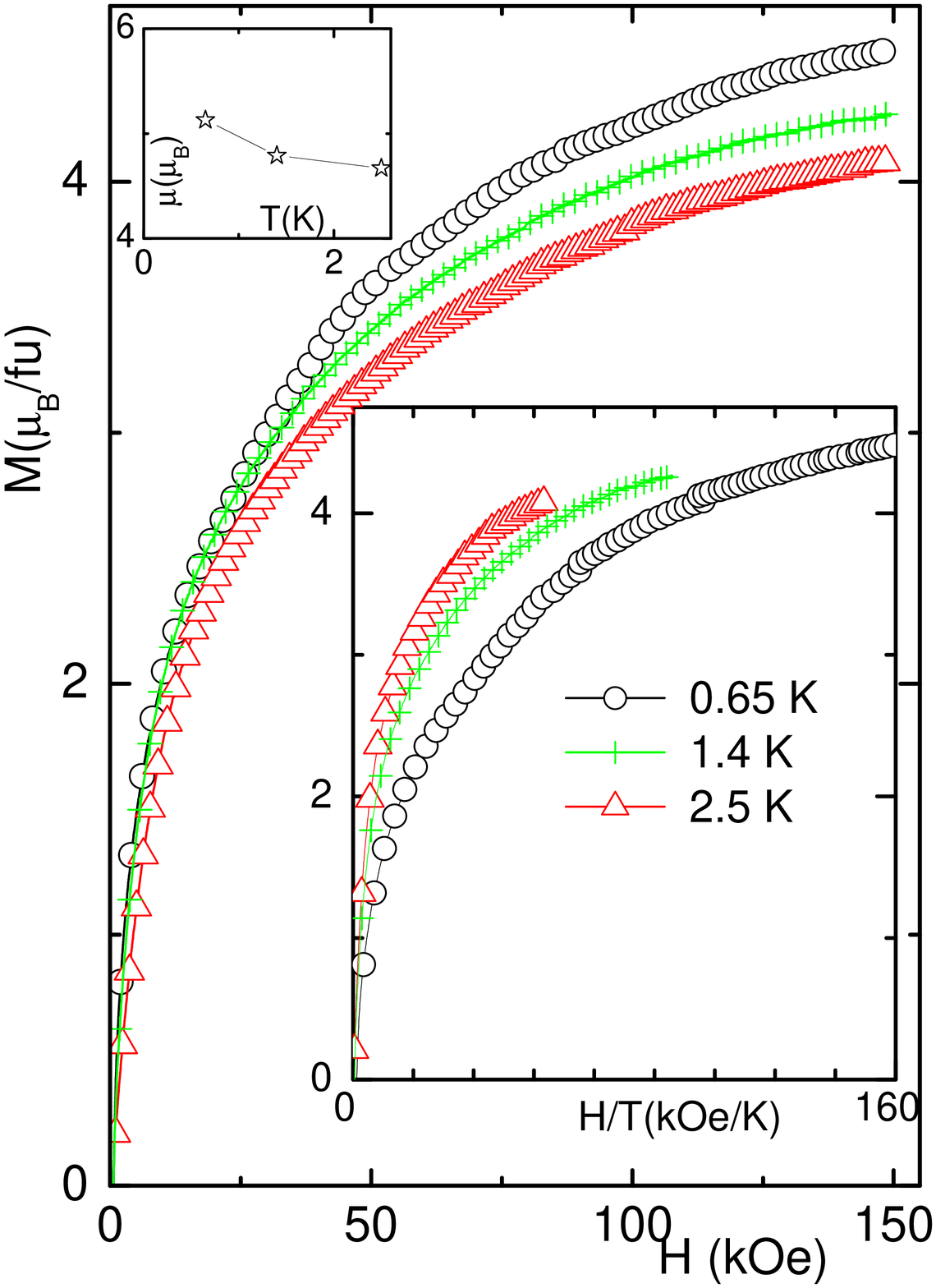}%
\caption{(Color online) Magnetization isotherms $M(H)$ of \textrm{TmCo}$_{2}%
$\textrm{B}$_{2}$\textrm{C} at different temperatures The lower-right inset
shows $M$ $versus$ $H/T$ curves while the upper-left inset shows the saturated
moment obtained from $\lim\limits_{1/H\rightarrow0}M(H)$. The lines are a
guide to the eye (see text).}%
\label{Fig.6}%
\end{center}
\end{figure}

\subsection{Magnetization}

Magnetization isotherms (Fig. \ref{Fig.6}) show that even for a field of 150
kOe, the magnetic moment per unit formula reaches only 4.5$\pm0.2$
$\mu_{\text{B}}$ ($T$=0.65 K) which is\ 65\% of the value expected for a free
\textrm{Tm}$^{3+}$ ion. Such a high-field susceptibility is attributed to the
magnetic response of the Co subsystem. The total moment saturation
is\ estimated from the extrapolation $\lim\limits_{1/H\rightarrow0}M(H)$: the
saturated moment (upper-left inset of Fig. \ref{Fig.6}) increases smoothly as
the temperature is decreased. On the other hand, the $M$ $versus$ $H/T$ curves
(lower-right inset of Fig. \ref{Fig.6}) reveal that the isotherms at 1.4 and
2.5 K almost collapse on each other but that of 0.65 K$\ $is distinctly
different. This observation supports our earlier arguments that \textrm{TmCo}%
$_{2}$\textrm{B}$_{2}$\textrm{C}\ orders magnetically with an onset point
lower than 1.4 K: within the paramagnetic state, all $M$($H/T)$ curves follow
the same Brillouin function and as such collapse on each other while below
$T_{C}$, the rapid development of the ordered moment ensures that each of the
$M$($H/T)$ isotherms is distinctly different from one another and, in
particular, from all paramagnetic isotherms.

\ Chang \textit{et al.}\cite{Chang96-TmNi2B2C-mag-struct} observed that the
moment of \textrm{TmNi}$_{2}$\textrm{B}$_{2}$\textrm{C}\ increases as the
temperature decreases; namely $\mu($1.2 K$)$ =3.74 $\mu_{\text{B}}$ and $\mu
($50 mK$)$ = 4.8 $\mu_{\text{B}}$. Comparing these features with those of Fig.
\ref{Fig.6}, it is inferred that the magnitude and thermal evolution of the
moments of both isotherms\ are similar which suggests a similarity in the
low-lying CEF level scheme. As such it is expected that the orientation of the
Tm moment to be along the $c$ axis just as observed in \textrm{TmNi}$_{2}%
$\textrm{B}$_{2}$\textrm{C}. The observed difference in the Tm$^{3+}$ moment
strength of the two isomorphs may be due to a slight variation in the CEF
effects. Rapid-quench influence on the site-occupation may lead to a reduction
or even a complete collapse of the Tm moment (just as argued by Mulder
\textit{et al.}\cite{Mulder96-TmNi2B2C,Mulder98-TmNi2B2C} to explain the
reduction of the Tm moment in \textrm{TmNi}$_{2}$\textrm{B}$_{2}$\textrm{C}).%
\begin{figure}
[th]
\begin{center}
\includegraphics[
height=3.4765in,
width=2.5391in
]%
{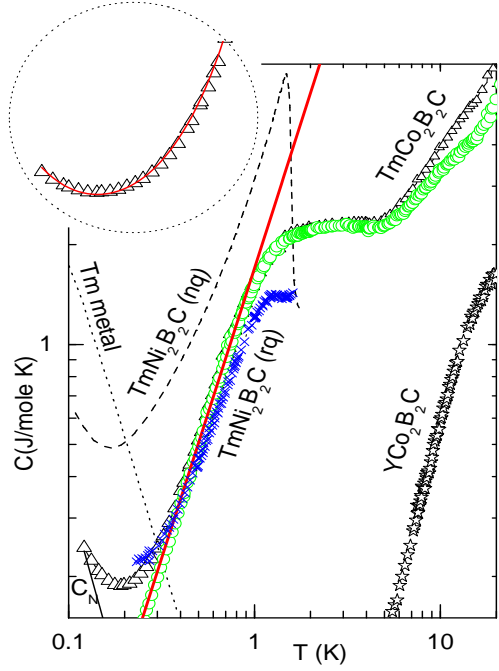}%
\caption{(Color online) The specific heat curve (triangles) of \textrm{TmCo}%
$_{2}$\textrm{B}$_{2}$\textrm{C}. For comparison, we include the specific heat
of \textrm{YCo}$_{2}$\textrm{B}$_{2}$\textrm{C}\ [stars, Ref. 16], the
non-quenched \textrm{TmNi}$_{2}$\textrm{B}$_{2}$\textrm{C} (dashed line, Ref.
15), the rapid-quench \textrm{TmNi}$_{2}$\textrm{B}$_{2}$\textrm{C} (cross,
Fig.1), and (only the nuclear contribution of ) the Tm-metal (dotted line,
Ref. 20). The magnetic contribution (circles) and the empirical function
$C_{M}(T)$=1.7$T^{\frac{7}{4}}$ J/moleK (thick solid line) are also indicated
(see text). The inset \ shows an expansion of the low-temperature tail of the
total specific heat of \ \textrm{TmCo}$_{2}$\textrm{B}$_{2}$\textrm{C
}(triangles) together with $C_{tot}$= $C_{ep}$(\textrm{YCo}$_{2}$%
\textrm{B}$_{2}$\textrm{C})+$0.035T^{-2}$+$1.7T^{\frac{7}{4}}$ J/molK (solid
line).}%
\label{Fig.7}%
\end{center}
\end{figure}

The low-field part of the $M$($H$, 0.65 K) curve of Fig. \ref{Fig.6} does not
manifest those characteristic features (such as convex curvature at a spin
flop event) which can be taken as indicative of an AFM\ ground state; instead
it shows a monotonic and steep increase which is typical of a forced
magnetization of a FM-like state. It is not uncommon in the magnetism of the
intermetallics that the character of the magnetic ground state is different
from the one suggested by the sign of the Curie-Weiss temperature.%
\begin{figure}
[th]
\begin{center}
\includegraphics[
height=3.0943in,
width=2.3289in
]%
{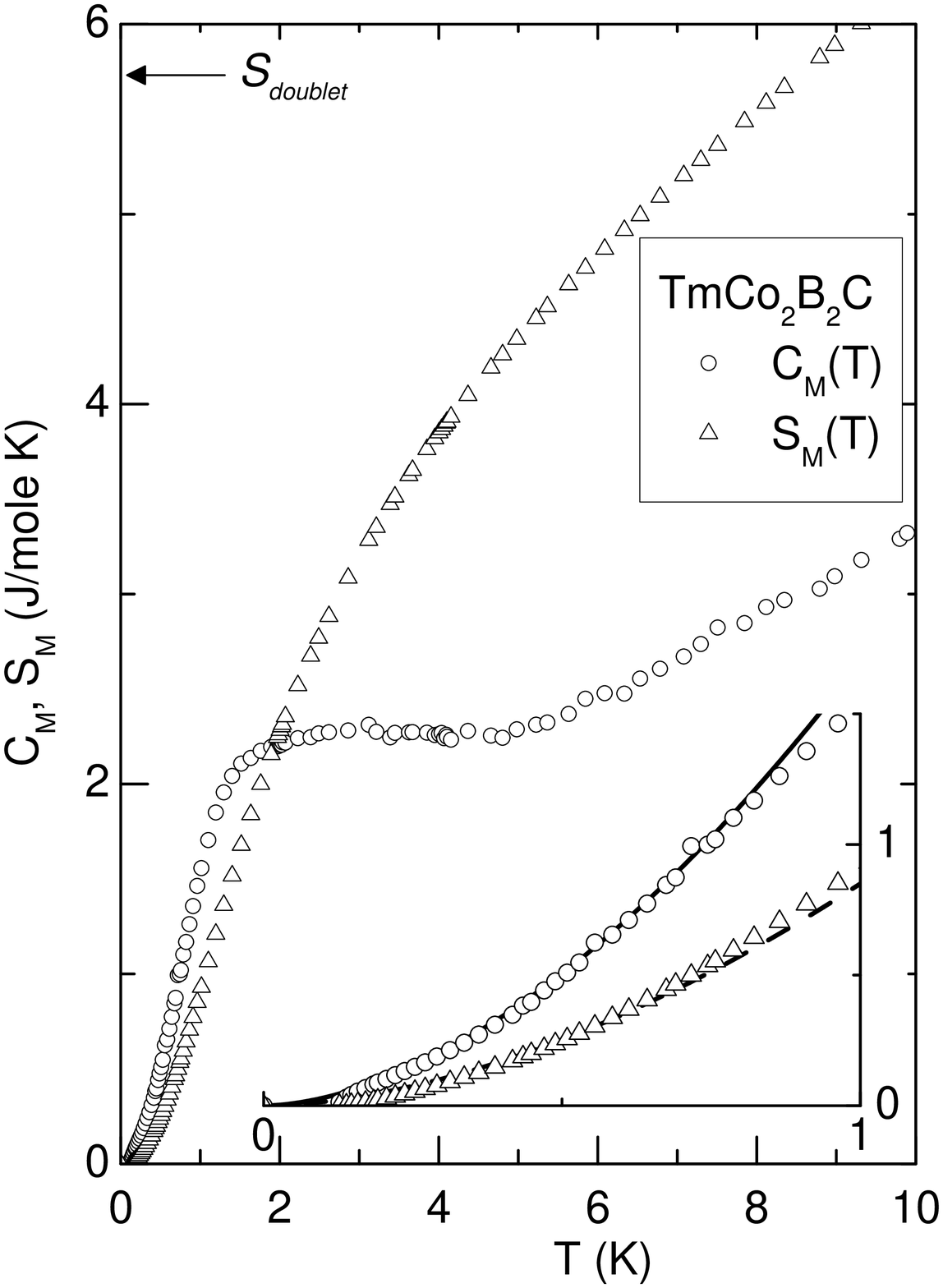}%
\caption{The magnetic contribution to the specific heat (circles) and magnetic
entropy (triangles) of \textrm{TmCo}$_{2}$\textrm{B}$_{2}$\textrm{C}. The
expected entropy for a doublet is shown as a horizontal arrow. The inset
shows, on an expanded scale, the experimental curves together with their
low-temperature fits: $C_{M}(T)$=1.7$T^{\frac{7}{4}}$ J/moleK (solid line) and
$S_{M}(T)$=0.85$T^{\frac{7}{4}}$ J/moleK (dashed line).}%
\label{Fig.8}%
\end{center}
\end{figure}

\subsection{Specific Heat}

The zero-field $C(T)$ curve of \textrm{TmCo}$_{2}$\textrm{B}$_{2}$\textrm{C}
is shown in Fig. \ref{Fig.7}. The low-temperature part of the total
$C(T)$\textrm{ }is largely due to the magnetic contribution, $C_{M}(T)$, which
is obtained after subtracting the contributions due to both the diamagnetic
reference (\textrm{YCo}$_{2}$\textrm{B}$_{2}$\textrm{C},
Ref.9\nocite{00-RCo2B2C}) and the nuclear interaction ($C_{N}\approx3.5T^{-2}$
mJ/mole K). It is interesting to note that the associated hyperfine
interaction (and the electronic magnetic moment) is much smaller than that of
Tm-metal ($C_{N}\approx26.8T^{-2}$ mJ/mole
K),\cite{Holmstrom-Nclr-Pr-Tm-Schottky} but closer, as expected, to that of
\textrm{TmNi}$_{2}$\textrm{B}$_{2}$\textrm{C}\ ($C_{N}\approx4.8T^{-2}$
mJ/mole K).\cite{Movshovich94-TmNi2B2C}%
\begin{figure}
[th]
\begin{center}
\includegraphics[
height=3.6391in,
width=2.5901in
]%
{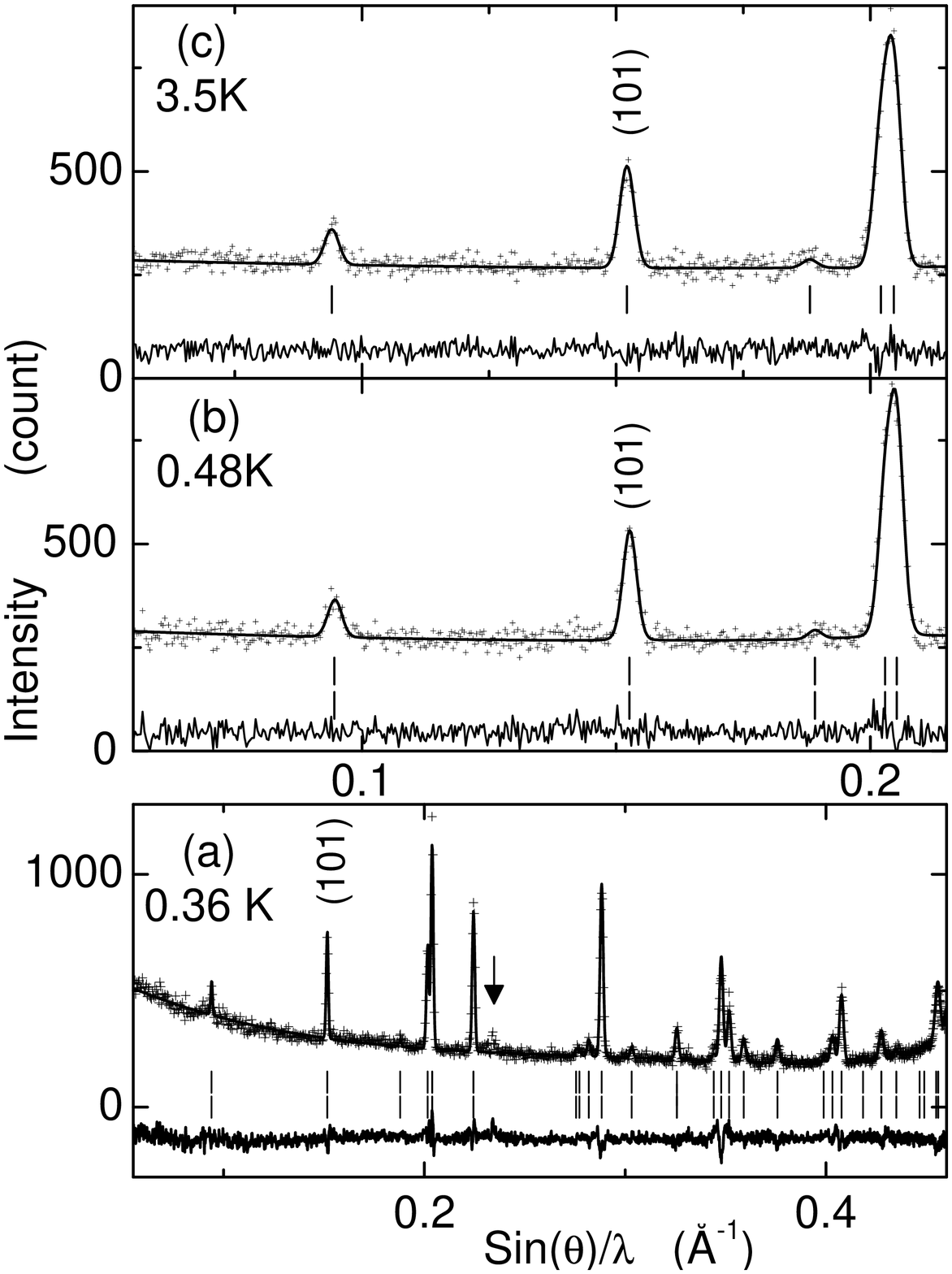}%
\caption{Neutron-diffractograms of \textrm{TmCo}$_{2}$\textrm{B}$_{2}%
$\textrm{C}, measured at (a) NIST high resolution diffractometer, $\lambda
$=2.0787 $\operatorname{\mathring{A}},$ and (b, c) NIST triple axis
diffractometer $\lambda$=2.359 $\operatorname{\mathring{A}}$. The small
vertical arrow indicates an impurity peak (see Figs. \ref{Fig.1} and
\ref{Fig.3} ). A noticeable increase in the low angle intensities can be
observed in lower panel but not in the upper panels: this suggests that this
broadening is instrument-dependent and not due to a small angle scattering
arising from magnetic or structural disorder induced by the rapid quench
process.}%
\label{Fig.9}%
\end{center}
\end{figure}

Figures \ref{Fig.7}-\ref{Fig.8} show that $C_{M}(T<0.8$ K$)$ =1.7$T^{\frac
{7}{4}}$ J/mole K; above 0.8 K it turns slowly into a plateaux which extends
up to 5 K; above this temperature it resumes the monotonic increase but with a
slower rate. This power-type thermal evolution of $C_{M}(T)$ (with an exponent
being close to $\frac{3}{2}$) is consistent with what one would expect from a
magnon contribution of a FM ordered state wherein the dispersion relation is
quadratic and the anisotropic field is vanishingly
small.\cite{03-Magnon-RNi2B2C,03-AFSup-RNi2B2C-Cm}

The magnetic entropy (Fig. \ref{Fig.8}) evolves as $S_{M}(T<0.8$
K$)$=0.85$T^{\frac{7}{4}}$ J/mole K, increases steadily for $T>$0.8 K, and
approximates the doublet-value $R\ln(2)$ only above 8 K. As all contributions
from the Co subsystem has already been subtracted, then this entropy
contribution must be due solely to the Tm subsystem and as such the
lowest-lying two levels of Tm$^{3+}$ are separated from the highest ones by an
energy gap which must be much higher than 8 K: indeed inelastic neutron
scattering on the isomorphous \textrm{TmNi}$_{2}$\textrm{B}$_{2}$\textrm{C}
showed that the doublet ground-state is separated from the first excited-state
by 30
K.\cite{Gasser97-CEF-RNi2B2C,Gasser96-CEF-HoErTmNi2B2C,rotter01-137,Gasser98-TmNi2B2C}
We have carried out preliminary inelastic neutron measurements on BT-7 to
search for crystal field excitations in \textrm{TmCo}$_{2}$\textrm{B}$_{2}%
$\textrm{C}. We did not find any sharp excitations up to an energy transfer of
20 meV, but we did observe at 4.0 K a distribution of scattering (not shown)
that appeared quasi-elastic, with the energy resolution of 1.0 meV employed.
The wave vector and temperature dependence of this scattering indicated that
it was magnetic in origin, and fitting with a Lorentzian distribution gave a
half width $\Gamma$= 3.9(4) meV. Further studies are underway.

As seen in Fig. \ref{Fig.2}(b), the$\ C(T)$\ curve of the $rq$ \textrm{TmNi}%
$_{2}$\textrm{B}$_{2}$\textrm{C} manifests the magnetic phase transition as a
broadened and weak magnetic event. Along the same line of reasoning, the
shoulder-like region in Figs. \ref{Fig.7}-\ref{Fig.8} is considered to be the
region within which the magnetic phase transition of \textrm{TmCo}$_{2}%
$\textrm{B}$_{2}$\textrm{C} sets-in (see the last comment in \S \ II.B). Then
the critical temperature is taken to be the point of the maximum slope which
is found to be 0.8 K, in excellent agreement with the considerations of Figs.
\ref{Fig.4}-\ref{Fig.6}.

%

\begin{figure}
[t]
\begin{center}
\includegraphics[
height=4.6882in,
width=3.5933in
]%
{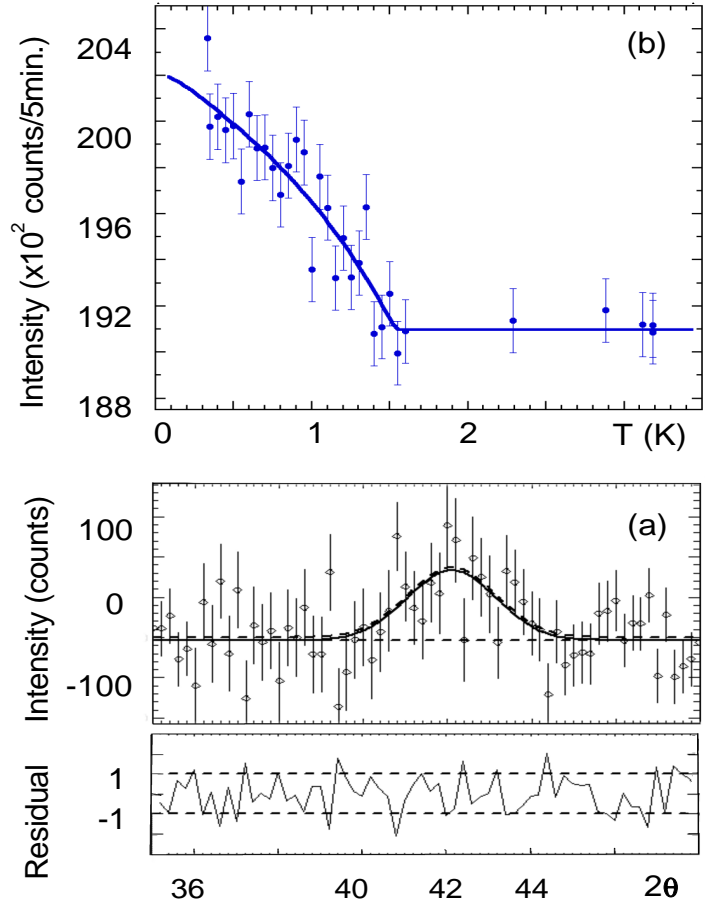}%
\caption{(a) The difference pattern of the (101) peak obtained after
subtracting the intensities measured at 4.0 K from those measured at 0.45 K.
The line is a Gaussian fit to the experimental points (the residual graph
represents the difference between the fit and the data). (b) Thermal evolution
of the intensity of the (101) Bragg peak. Each point represents a measurement
of the peak using a single setting of the Position Sensitive Detector on BT-7.
The solid curve is a fit to mean field theory; this gives a critical
temperature of 1.54(8) K. Error bars in this figure are statistical and
represent one standard deviation. }%
\label{Fig.10}%
\end{center}
\end{figure}

\subsection{Neutron diffraction}

The high-resolution neutron diffractograms (Fig. \ref{Fig.9}) have been
measured on BT-1 down to 0.36 K using a He3 cryostat and for as low angle as
possible. Within the measured temperature range and experimental accuracy, no
additional Bragg peaks that might be associated with magnetic order are
observed. We therefore carried out diffraction measurements using the
high-intensity/coarse resolution thermal triple axis instruments BT-9 and BT-7
to search for magnetic order. Within the angular range up to 65 degrees in
scattering angle, the only significant change in intensity was observed for
the (101) Bragg peak, as shown by the difference plot in Fig. \ref{Fig.10}
(a). The temperature dependence of the intensity of this peak is shown in Fig.
\ref{Fig.10} (b), and the increase at lower temperatures indicates that
magnetic order has developed. The solid curve is a fit of mean field theory to
the data, which indicates that the magnetic ordering temperature is 1.54 K;
this critical point (though determined from mean field analysis which is not
well suited for such a determination) is nearly twice the value estimated from
the susceptibility (Fig. \ref{Fig.5}) and the specific heat (Fig.
\ref{Fig.7}). Considering that the $rq$ process is observed to induce a
drastic broadening in the transition region (see Fig. \ref{Fig.2}), then it is
most probable that the difference in the values of the transition point is due
to the fact these techniques are differing in their frequency window.

The observation of a magnetic contribution at the (101) peak suggests that the
magnetic wave vector is (000). It is emphasized that the weak impurities do
not influence this analysis since we used the subtraction method (see Fig.
\ref{Fig.10}). Below we argue that the only magnetic structure which is
compatible with this (000) wave vector is the FM state. Such a conclusion is
consistent with the above mentioned arguments about the forced magnetization
(Fig. \ref{Fig.6}) and the magnon contribution of the specific heat (Fig.
\ref{Fig.7}). As the crystal structure is body centered tetragonal (no
orthorhombic distortion is expected since magnetostriction effects are weak)
with only one Tm atom per primitive cell, then no order of the Tm sublattice
can be antiferromagnetic.\ However, there are two Co atoms per primitive cell
and as such an AFM\ order may develop but this should be discarded since there
are no magnetic moments on the Co sublattice (see above). Thus it is concluded
that this magnetic order must be due to the FM order of the Tm sublattice: it
is recalled that both the saturated and effective moments are in excellent
agreement with those found for \textrm{TmNi}$_{2}$\textrm{B}$_{2}$\textrm{C}.
It is worth adding that our recent studies show that the manifestation of a FM
state in \textrm{TmCo}$_{2}$\textrm{B}$_{2}$\textrm{C} is not unusual: a\ FM
state is common among the magnetic $R$\textrm{Co}$_{2}$\textrm{B}$_{2}%
$\textrm{C} compounds ($R$=Dy, Ho, Er)\cite{08-Mag-Structure-RCo2B2C} and
\textrm{TbCo}$_{2}$\textrm{B}$_{2}$\textrm{C}.\cite{08-Mag-Structure-TbCo2B2C}

On comparing the obtained results of \textrm{TmCo}$_{2}$\textrm{B}$_{2}%
$\textrm{C} with those of \textrm{TmNi}$_{2}$\textrm{B}$_{2}$\textrm{C}, it is
evident that their magnetic ground state are different: the former manifests a
colinear FM state while the later orders into a modulated spin density wave.
This difference is attributed to the difference in the indirect exchange
couplings which in turn is due to the difference in the electronic structures.
It is recalled that there is a clear difference between the electronic
structure of \textrm{LuNi}$_{2}$\textrm{B}$_{2}$\textrm{C} and \textrm{LuCo}%
$_{2}$\textrm{B}$_{2}$\textrm{C}%
.\cite{Coehoorn94-RNi2B2C-electronic-structure} On the other hand, the
observation that their magnetic moments are similar is an indication that the
CEF properties of the Tm$^{3+}$ ions have not been greatly modified by the
$rq$ process or by the interchange of the 3$d$ atoms.

\textrm{TmCo}$_{2}$\textrm{B}$_{2}$\textrm{C} shows no trace of
superconductivity in\textrm{ }the $ac$ -susceptibility (down to 20 mK), the
specific-heat (down to 100 mK), or the preliminary resistivity curves measured
down to 20 mK (not shown). It is recalled that none of the reported members of
the $R$\textrm{Co}$_{2}$\textrm{B}$_{2}$\textrm{C} series is a
superconductor,\cite{00-RCo2B2C} not even \textrm{YCo}$_{2}$\textrm{B}$_{2}%
$\textrm{C} in spite of having a Sommerfeld constant and Debye temperature
that are almost equal to those of superconducting \textrm{YNi}$_{2}$%
\textrm{B}$_{2}$\textrm{C}; \textrm{YCo}$_{2}$\textrm{B}$_{2}$\textrm{C}
manifests spin fluctuation features,\cite{04-Pr(CoNi)2B2C} a property assumed
to be unfavorable to the onset of the superconductivity. These spin
fluctuation (associated with the Co sublattice) together with the presence of
the FM state (associated with the Tm sublattice) are detrimental to the
presence of the superconductivity in \textrm{TmCo}$_{2}$\textrm{B}$_{2}%
$\textrm{C}.

\section{Conclusion}

A single-phase \textrm{TmCo}$_{2}$\textrm{B}$_{2}$\textrm{C }has been
successfully stabilized via the rapid-quench process. Its crystal structure is
isomorphous to that of the body-centered tetragonal \textrm{TmNi}$_{2}%
$\textrm{B}$_{2}$\textrm{C}.\ The paramagnetic properties are characterized by
a modified CW behavior while the low-temperature features are dominated by an
ordering of\ Tm$^{3+}$ moments. Based on the character of the low-temperature
magnon contribution to the specific heat, the characteristic feature of the
low-temperature forced magnetization, and the analysis of the magnetic
contribution to the neutron diffractograms, it is inferred that the magnetic
order of the Tm sublattice is FM. At 0.65 K and under an applied magnetic
field of 150 kOe, the Tm$^{3+}$ moment is observed to reach 4.5$\pm0.2$
$\mu_{\text{B}}$; though only 65\% of the value expected for a free
\textrm{Tm}$^{3+}$ ion, this value is only 6\% lower than the moment of
\textrm{TmNi}$_{2}$\textrm{B}$_{2}$\textrm{C}: it is concluded then that the
CEF effects in both isomorphs are similar. Finally, no superconductivity is
observed in \textrm{TmCo}$_{2}$\textrm{B}$_{2}$\textrm{C }down to 20 mK: a
feature attributed to the presence of the spin fluctuation in the Co subsystem
and the FM ordering of the Tm sublattice..

\begin{acknowledgments}
We acknowledge the partial financial support from the Brazilian agencies CNPq
(485058/2006-5) and Faperj (E-26/171.343/2005). Identification of commercial
equipment in the text is not intended to imply recommendation or endorsement
by National Institute of Standards and Technology. We would like to thank
Jiying Li for his assistance with the data analysis using the position
sensitive detector on BT-7.
\end{acknowledgments}

\bibliographystyle{apsrev}
\bibliography{Borocarbides,Intermetallic,Mag-classic,Massalami,Nuclear-hyperfine-MES,Sup-classic,sup-Mag-interplay,To-Be-Published}

\end{document}